\documentclass[10pt,letterpaper,twocolumn]{article} %% two column, final layout

\usepackage{ol2}
\usepackage[draft]{hyperref}
\usepackage{amsmath}

\begin{document}

\twocolumn[ %% activate for two-column option

\title{$\mathcal{PT}$-symmetric microring laser-absorber}

%% For REVTeX it is possible to automate superscript and e-mail callouts with the superscriptaddress option; see REVTeX4 documentation.

\author{Stefano Longhi}

\address{Dipartimento di Fisica, Politecnico di Milano and Istituto di Fotonica e Nanotecnologie del Consiglio Nazionale delle Ricerche, Piazza L. da Vinci 32, I-20133 Milano, Italy (stefano.longhi@polimi.it)}

\author{Liang Feng}

\address{Department of Electrical Engineering, The State University of New York at Buffalo, Buffalo, NY 14260, USA (fengl@buffalo.edu)}

\begin{abstract}
The lasing and coherent perfect absorption (CPA) properties of $\mathcal{PT}$-symmetric microrings with mixed index and gain gratings, externally coupled to a bus waveguide, are theoretically investigated. For a complex grating at the $\mathcal{PT}$ symmetry breaking point, perfect unidirectional (either clockwise or counterclockwise) laser emission can be realized, however the grating does not discriminate longitudinal modes and CPA can not be simultaneously achieved. Above the  grating  $\mathcal{PT}$ symmetry breaking point, single mode emission and simultaneous CPA can be obtained, with unbalanced and controllable excitation of clockwise and counterclockwise modes in the ring.   
\end{abstract} 


%(130.0130) Integrated optics; 
%(140.0140) Lasers and laser optics; 
%(140.3490) Lasers, distributed-feedback

 ] %% activate for two-column option
{\it Introduction.} Optical systems that combine balanced loss and gain profiles offer a versatile platform to study the physics of parity-time ($\mathcal{PT}$) symmetric Hamiltonians \cite{Bender}, featuring new synthetic materials with novel properties \cite{g1,g2,g3}. In this context, it has been shown \cite{g4,g5,g6,g6bis} that an active structure satisfying the  $\mathcal{PT}$ symmetry condition $\epsilon(-\mathbf{r})=
\epsilon^*(\mathbf{r})$ can behave simultaneously as a laser oscillator and as a coherent perfect absorber (CPA),  i.e., it
can simultaneously emit coherent outgoing waves and fully absorb incoming coherent waves with appropriate amplitudes and phases. 
This has lead to the introduction of the so-called $\mathcal{PT}$-symmetric laser-absorber devices \cite{g4}, which have been demonstrated in recent works at microwaves \cite{referee1,referee2}.
Distributed-feedback structures (DFB) with combined index and gain/loss gratings \cite{uff1,uff2,uff3,uff4} provide a natural platform to study $\mathcal{PT}$-symmetric Hamiltonians and laser-absorber systems in the optical region \cite{g4,g6bis,g7,g8,g9,g10}. At the full quantum level, $\mathcal{PT}$-symmetric lasers show interesting lasing resonance line shapes and quantum noise properties \cite{Schom1,Schom2}.
Integrated semiconductor microring cavities have provided over more than one decade attractive devices for applications to wavelength filtering, dispersion compensation, optical switching and optical memories (see, e.g.,  \cite{semi,memo,nature}). In a series of recent works \cite{teo,exp1,exp2,exp3}, semiconductor microring/microdisk lasers in $\mathcal{PT}$ symmetric configurations have been introduced and realized, showing nonreciprocal transmission, optical isolation and mode selection. In particular, single-mode emission from a $\mathcal{PT}$-synthetic DFB microring laser with a pure gain grating has been realized by Feng and collaborators \cite{exp3}. However, the demonstration of a $\mathcal{PT}$-synthetic laser-absorber device at optical frequencies is still missing.\\
In this Letter we consider a class of $\mathcal{PT}$-synthetic microring lasers with mixed gain and index gratings, externally coupled to a bus waveguide, and investigate their lasing and CPA properties. It is shown that, for a mixed grating at the $\mathcal{PT}$ symmetry breaking point, perfect unidirectional (either clockwise or counterclockwise) laser emission is obtained, however simultaneous CPA and longitudinal mode discrimination can not be realized in this case. Conversely, for a mixed grating in the broken $\mathcal{PT}$ phase single mode emission and simultaneous CPA can be obtained, with unbalanced and controllable excitation of clockwise and counterclockwise waves in the ring. 
\par
{\it $\mathcal{PT}$-symmetric microring lasers with mixed gratings.}  Let us consider a circular microring/microdisk laser of radius $R$ side coupled to a bus waveguide in the geometrical setting shown in Fig.1(a). Indicating by $a$, $b$ and by $c$, $d$ the amplitudes of the outgoing and incoming light fields at frequency $\omega$ in the bus waveguide, respectively, the CPA and lasing properties of the coupled microring-waveguide system can be derived from the analysis of the $ 2 \times 2$ transfer matrix $\mathcal{Q}=\mathcal{Q}(\omega)$ defined by
\begin{equation}
\left( 
\begin{array}{c}
a \\
d
\end{array}
\right)= \mathcal{Q}(\omega) 
\left( 
\begin{array}{c}
c \\
b
\end{array}
\right).
\end{equation}
In particular \cite{g4,g7}: (i) the system is below lasing threshold if all the zeros of $\mathcal{Q}_{22}(\omega)$ lie in lower half complex plane $\rm{Im}(\omega)<0$; the most unstable (lasing) mode as some control parameter of the system is varied corresponds to the first zero of $Q_{22}$ that crosses the real axis from below at the frequency $\omega_L$; correspondingly, the amplitudes of outgoing waves in the bus waveguide are related by the relation $a/b=\mathcal{Q}_{12}(\omega_L)$; (ii) CPA occurs
if $\mathcal{Q}_{11}(\omega)=0$ at some frequency $\omega=\omega_0$ on the real axis and for the appropriate coherent excitation of the system $c=-\mathcal{Q}_{12}(\omega_0) d$; (iii) for a $\mathcal{PT}$ invariant system, one has $\mathcal{Q}_{11}(\omega)=\mathcal{Q}_{22}^*(\omega^*)$, and thus CPA and lasing can occur simultaneously ($\omega_L=\omega_0$); (iv) while the system may show different reflections for left and right incidence sides, reciprocity is not violated and the transmissions from left and right incidence sides are the same.\\
Simple configurations of active/passive microrings coupled to a bus waveguide have been investigated in several previous works; for example, the CPA property of a lossy microring without any grating that occurs at a certain waveguide-microring coupling (known as {\it critical coupling}) was studied in Refs.\cite{Yariv,Vahala}. Here we extend such previous studies by considering a microring with a mixed index and gain grating in a $\mathcal{PT}$-symmetric configuration \cite{g8,g9}, highlighting some interesting operational regimes. To this aim, let us assume that the effective refractive index $n(s)$ along the curvilinear coordinate $s$ of the ring is weakly modulated, with periodicity $\Lambda= \pi / k_B$,  according to $n(s)=n_0+i \sigma +\delta n_R \cos(2 k_B s)-i \delta n_I \sin (2 k_B s)$, where $\delta n_{R,I}$ are the modulation depth of index and gain/loss gratings, respectively, $n_0$ is the substrate index of the medium, and $\sigma$ describes the effective mean loss/gain term. Note that a non-vanishing value of $\sigma$ implies that the system is not $\mathcal{PT}$ invariant. The electric field $\mathcal{E}$ at frequency $\omega$ close to the Bragg frequency $\omega_B \equiv c_0 k_B / n_0$ circulating in the microring can be written as a superposition of counterclockwise and clockwise waves of amplitudes $u(s)$ and $v(s)$, respectively, namely $\mathcal{E}(s,t)=u(s) \exp(ik_B s-i \omega t)+v(s) \exp(-ik_Bs-i \omega t) +c.c.$ , where the amplitudes $u$ and $v$ satisfy the coupled-mode equations \cite{uff1,g9}
\begin{equation}
i \frac{du}{ds}=-\delta u - \rho_1 v \; ,\;\; i \frac{dv}{ds}=\delta v + \rho_2 u
\end{equation}
where $\delta = \delta_R+i \gamma$, $ \delta_R=(n_0/c_0)(\omega-\omega_B)$ is the normalized frequency detuning parameter, $\gamma=(k_B / n_0) \sigma$ is the mean loss rate per unit length (gain rate if $\gamma<0$), and $\rho_1=k_B (\delta n_R-\delta n_I)/(2n_0)$, $\rho_2=k_B (\delta n_R+\delta n_I)/(2n_0)$ are the asymmetric coupling rates of counter-propagating waves. The length $L=2 \pi R$ of the ring is assumed to be an integer multiple of the grating period $\Lambda$. The amplitudes $u$, $v$ of counter-propagating waves at $s=0^+$ and $s=L^+$ are related by the relation 
\begin{figure}[htb]
\centerline{\includegraphics[width=8.3cm]{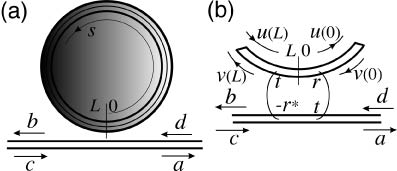}} \caption{
\footnotesize {(Color online)  (a) Schematic of  a microring side-coupled to a bus waveguide; (b) magnification of the coupling region.}}
\end{figure}
\begin{equation}
\left( 
\begin{array}{c}
u(L^-) \\
v(L^-)
\end{array}
\right)= \mathcal{M}(\omega) 
\left( 
\begin{array}{c}
u(0^+) \\
v(0^+)
\end{array}
\right),
\end{equation}
where the transfer matrix $\mathcal{M}$ is obtained by solving coupled-mode equations (1) and reads
\begin{equation}
\mathcal{M}=\left(
\begin{array}{cc}
\cos( \lambda L)+i \frac{\delta}{\lambda} \sin( \lambda L) & i \frac{\rho_1}{\lambda} \sin (\lambda L) \\
-i \frac{\rho_2}{\lambda} \sin (\lambda L) & \cos (\lambda L)-i \frac{\delta}{\lambda} \sin (\lambda L)
\end{array}
\right)
\end{equation}
where $\lambda \equiv \sqrt{\delta^2-\rho_1 \rho_2}$. Let us first consider the case of an ideal lossless microring ($\gamma=0$) uncoupled to the bus waveguide. This 
case differs from the unidirectional invisibility system of Refs.\cite{g8,g9} because of the ring  periodic boundary conditions $u(L^-)=u(0^+)$ and $v(L^-)=v(0^+)$, which using Eq.(4) yields the condition $\cos(\lambda L)=1$, i.e. $\delta= \pm \sqrt{(2 \pi l/L)^2+\rho_1 \rho_2}$, where $l= 0, \pm1, \pm2, ...$ is the longitudinal mode index. Such a relation shows that unstable modes arise for $\rho_1 \rho_2  \leq0$, i.e. $\mathcal{PT}$ symmetry breaking is attained  at $\rho_1=0$ ($\delta n_I=\delta n_R$). Interestingly, such a symmetry breaking condition is the same than that obtained for unidirectional invisibility problem in Refs.\cite{g8,g9}. Coupling of the microring with the bus waveguide is described by a unitary matrix $\mathcal{T}$ [see Fig.1(b)]
\begin{figure}[htb]
\centerline{\includegraphics[width=8.3cm]{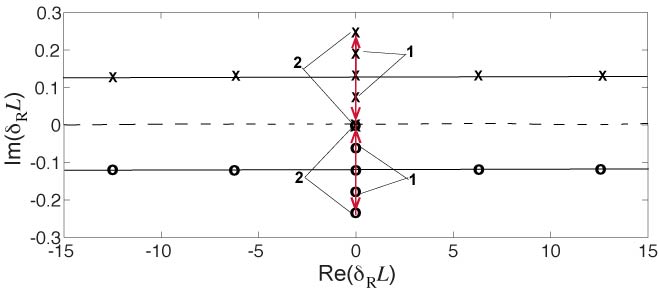}} 
\caption{ \footnotesize {(Color online) Trajectories of the zeros of $\mathcal{Q}_{22}(\omega)$ (open circles) and of $\mathcal{Q}_{11}(\omega)$ (crosses) in the complex frequency plane $\delta_RL=(n_0L/c_0)(\omega-\omega_B)$ for $t^2=0.8$, $\rho_2L=0.5$ and for increasing values of $|\rho_1L|$ (i.e. of $\delta n_I$). At $\rho_1L=0$, corresponding to $\delta n_I=\delta n_R$, the zeros of $\mathcal{Q}_{11}$ lie on the horizontal line ${\rm Im}(\delta_R L) ={\rm ln} \; t$ and the laser is below threshold. As $|\rho_1|L$ is increased, the zero with  ${\rm {Re}} (\delta_RL)=0$ (i.e. the mode with longitudinal index $l=0$) splits and the two splitted zeros move in opposite directions along the imaginary axis (see the arrows in the figure):  point 1 $\rho_1L=-0.0124$, point 2 $\rho_1L=-0.0248$, corresponding to the laser threshold. The splitting of the other zeros is extremely small and not visible in the figure. The zeros of $\mathcal{Q}_{11}$ are the complex conjugates than the zeros of $\mathcal{Q}_{22}$ and undergo specular trajectories.}}
\end{figure}
\begin{equation}
\left( 
\begin{array}{c}
u(0^+) \\
a
\end{array}
\right)= \mathcal{T}
\left( 
\begin{array}{c}
u(L^-) \\
c
\end{array}
\right) , \;  \left( 
\begin{array}{c}
v(L^-) \\
b
\end{array}
\right)= \mathcal{T}
\left( 
\begin{array}{c}
v(0^+) \\
d
\end{array}
\right),
\end{equation}
which we assume of the form \cite{Yariv,Vahala} 
\begin{equation}
\mathcal{T}=\left( 
\begin{array}{cc}
t & r \\
-r^* & t
\end{array}
\right),
\end{equation}
 where $t$ is the transmission coefficient through the coupler ($0 < t  <1$), and $r=i \sqrt{1-t^2}$. The chosen symmetry of the coupling matrix, which holds for a lossless coupler and without propagative phase shift, is appropriate for typical experimental ring-waveguide couplings \cite{Vahala}
and conserves the $\mathcal{PT}$ symmetric of the scattering problem. 
From Eqs.(3) and (5)  one readily obtains the following expressions for the elements of the transfer matrix $\mathcal{Q}$
\begin{eqnarray}
\mathcal{Q}_{11} & = & \mathcal{T}_{22} \mathcal{R} \left( 2-\mathcal{M}_{11}/ \mathcal{T}_{22}-\mathcal{M}_{22} \mathcal{T}_{22}   \right) \nonumber \\
\mathcal{Q}_{12} & =- & \mathcal{T}_{12}  \mathcal{T}_{21}  \mathcal{M}_{12} \mathcal{R} \; , \; \mathcal{Q}_{21} =-  \mathcal{T}_{12}  \mathcal{T}_{21}  \mathcal{M}_{21} \mathcal{R} \;\;\;\;\;\;\; \\
\mathcal{Q}_{22} & = & \mathcal{T}_{11} \mathcal{R} \left( 2-\mathcal{M}_{22}/ \mathcal{T}_{11}-\mathcal{M}_{11} \mathcal{T}_{11}   \right) \nonumber 
\end{eqnarray}
where $\mathcal{R}=(1-\mathcal{M}_{11}\mathcal{T}_{11}-\mathcal{M}_{22}\mathcal{T}_{22}+\mathcal{T}_{11}\mathcal{T}_{22})^{-1}$. Note that, for $\gamma=0$ $\mathcal{PT}$ invariance of the system is conserved and one has $\mathcal{Q}_{11}(\omega)=\mathcal{Q}_{22}^{*}(\omega^*)$. Note also that the 'cavityless'  scattering problem previously investigated in Ref.\cite{g8} in the regime of unidirectional invisibility is obtained from our analysis in the limiting case $t \rightarrow 0$.\\
{\it Unidirectional laser emission}.
Let us first consider the case of a mixed grating at the $\mathcal{PT}$ symmetry breaking point, i.e. $\delta n_I=\delta n_R$ corresponding to $\rho_1=0$. In this case the zeros of $\mathcal{Q}_{11}$ and $\mathcal{Q}_{22}$ in the  complex $\delta_R$ plane can be readily calculated and read
\begin{equation}
\delta_R=-i \left( \gamma \pm {\rm ln} (t) /L \right)+ 2 \pi l / L
\end{equation}
where $l=0 \pm1, \pm2, ...$ is the longitudinal mode index, and where the upper (lower) sign on the right hand side in Eq.(8) applies to $\mathcal{Q}_{11}$ ($\mathcal{Q}_{22}$). Note that, if the mean value of loss/gain $\gamma$ in the ring vanishes, the zeros of $\mathcal{Q}_{22}$ ($\mathcal{Q}_{11}$) lie in the $\rm{Im}( \omega) <0$ ($\rm{Im}( \omega) >0$) complex plane [see Fig.2(a)], regardless of the grating strength $\delta n_R=\delta n_I$, so that lasing and CPA can not be reached. However, if we break the $\mathcal{PT}$ invariance of the system by the introduction of a net mean loss or gain $\gamma$ in the ring, from Eq.(8) it follows that either lasing or CPA can be attained at $\gamma= \pm {\rm{ln}}(t) /L$, respectively, for all longitudinal modes. The CPA regime basically reproduces the critical coupling scheme of Refs.\cite{Yariv,Vahala} and, since $\mathcal{Q}_{12}=0$, it  requires a single input beam; the main difference here is that CPA is obtained solely for one side of the input signal that excites the clockwise mode $v$, namely $c=0, d \neq0$. The reason thereof is that, under such an input excitation, the grating appears to be invisible and Bragg scattering does not occur ($u=0$), thus our device behaves exactly in the scheme of Refs.\cite{Yariv,Vahala}. Conversely, in the other excitation direction ($d=0, c \neq 0$) Bragg scattering occurs and perfect absorption is not anymore possible.  More interestingly is the case of lasing, which requires a net gain $\gamma={\rm{ln}}(t) /L$ in the ring. In such a case because of $\mathcal{Q}_{12}=0$ one has $a=0$, i.e. laser oscillation is {\it unidirectional} (only the clockwise wave $v$ circulates in the ring, whereas $u=0$). Hence using a mixed gain/index grating with $\delta n_R= \delta n_I$ enables to obtain unidirectional laser emission, preventing the oscillation of the counterclockwise mode. This is a rather interesting result, since unidirectional oscillation in the ring does not require the introduction of non-reciprocal elements, e.g. magnetic or nonlinear elements \cite{Ramezani,referee3}. Besides unidirectionality, single longitudinal mode operation can be obtained for a small ring radius owing to the finite bandwidth of the gain medium.\\ 
{\it Mode selection and laser-absorber modes}. Let us consider the complex grating with $\gamma=0$ and in the broken $\mathcal{PT}$ phase ($\delta n_I> \delta n_R$). Figure 2 shows a typical behavior of the trajectories of the zeros of $\mathcal{Q}_{22}$ and $\mathcal{Q}_{11}$ in the complex $\omega$ plane as $\delta n_I$ is increased above $\delta n_R$ (note that, since $\gamma=0$, the zeros of $\mathcal{Q}_{11}$ are just the complex conjugates than the zeros of $\mathcal{Q}_{22}$).  At $\delta n_I= \delta n_R$, according to Eq.(8) all the zeros of $\mathcal{Q}_{22}$ lie in the ${\rm {Im}}(\delta_R)<0$ sector of the complex plane. As $\delta n_I$ is increased above $\delta n_R$, the central zero, corresponding to ${\rm Re}(\omega)=\omega_B$ (i.e.to the mode index $l=0$), is splitted into two zeros which move apart each other from opposite directions along the vertical (imaginary) axis as $\delta n_I$ is increased. The other zeros of $\mathcal{Q}_{22}$, corresponding to modes with $l \neq 0$, are almost unchanged, undergoing a small splitting in the horizontal (real axis) direction which is not visible in the scale of the figure. Laser threshold and simultaneous CPA occur when one zero of $\mathcal{Q}_{22}$ and one of $\mathcal{Q}_{11}$ touch on the real axis (point 2 in Fig.2). This occurs when $\cos ( \sqrt{- \rho_1 \rho_2}L)=2/(t+1/t)$, corresponding to the critical value of $\delta n_I$
\begin{equation}
\delta n_I^2=\delta n_R^2+ \left( \frac{2 n_0}{k_B L} \right)^2 {\rm {acos}}^2 \left( \frac{2t}{1+t^2}\right).
\end{equation}
\begin{figure}[htb]
\centerline{\includegraphics[width=8.3cm]{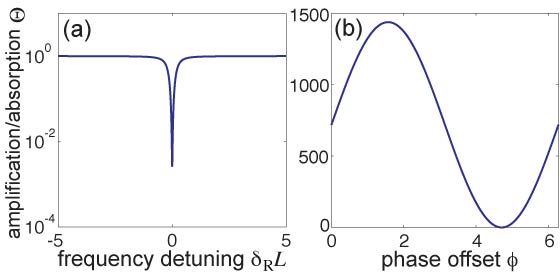}} 
\caption{ \footnotesize {(Color online)  Behavior of the overall absorption/amplification coefficient $\Theta$ for two-beam input excitation, defined by Eq.(10) given in the text, versus (a) the normalized frequency detuning parameter $\delta_R L$, and (b) the relative phase $\phi$ of the two input beams. Parameter values are given in the text.}}
\end{figure}

In this case, single-mode operation is expected at lasing threshold. For the lasing mode one has $|a/b|^2=|\mathcal{Q}_{12}|^2=| \rho_1 / \rho_2|$, whereas for the CPA mode one has $|c/d|^2=|\mathcal{Q}_{12}|^2=| \rho_1 / \rho_2|$. Note that, for $\delta n_R=0$, i.e. for a pure gain grating, $\rho_2=-\rho_1$ and thus $|a/b|^2=1$, i.e. laser emission occurs symmetrically from both waveguide ends and the two clockwise and counter-clockwise waves in the ring are equally excited. This kind of mode selection with balanced excitation of counterporpagating waves has been recently demonstrated in Ref.\cite{exp3}. Similarly, CPA requires balanced intensities of the two exciting beams. The introduction of the index grating, in addition to the gain grating, makes the laser emission (and similarly the CPA mode) asymmetric. In particular, a nearly unidirectional laser emission occurs if $|\rho_1 / \rho_2| \gg 1$, which is obtained in the low-coupling limit $t \rightarrow 1^-$.  For example, for $t^2=0.8$ and $\rho_2L=0.5$, the lasing threshold (and the CPA regime) is attained at $\rho_1L=-0.0248$, with an almost unidirectional laser emission ($|a/b|^2 \simeq 0.05$).Hence the complex grating in the broken $\mathcal{PT}$ phase $\delta n_I > \delta n_R$ can operate simultaneously as a laser and as a CPA device, providing single mode selection and enabling to control the unidirectional/bidirectional emission ratio by tuning of the index/gain grating strengths. To highlight the simultaneous amplifying/absorbing properties of the device in the $\delta n_I > \delta n_R$ and $\gamma=0$ regime, we numerically computed the  overall absorption/amplification coefficient $\Theta(\omega)$ of the structure for coherent two-beam excitation just below the lasing threshold.  The coefficient $\Theta(\omega)$ is defined as the ratio of the total intensity of outgoing (reflected or transmitted) waves to the total intensity of the incoming (injected) waves \cite{g4} in the bus waveguide, that is,
\begin{equation}
\Theta(\omega)=(|a|^2+|b|^2)/(|c|^2+|d|^2).
\end{equation}
Note that the vanishing of $\Theta$ is the signature of CPA, whereas $|\Theta|>1$ indicates that an overall amplification has been realized in the system. Figure 3(a) shows the behavior (on a logarithmic scale) of $\Theta$ versus the normalized frequency detuning $\delta_RL=(n_0 L/c_0)(\omega-\omega_B)$ for a two-beam excitation with amplitudes $c$ and $d$ satisfying the condition $c/d=-\mathcal{Q}_{12}(\omega_B) \simeq 0.211i$ and for parameter values $t^2=0.8$, $\rho_2 L=0.5$ and $\rho_1L=-0.0223$. Such a value of $\rho_1L$ is $10 \%$ below the threshold value for laser oscillation [see Fig.2(c)]. Note that, as expected, at $\omega= \omega_B$ the incident light is almost fully absorbed, which is the signature of CPA. Figure 3(b) shows the behavior of $\Theta$ for coherent input excitation $c/d=|\mathcal{Q}_{12}(\omega_B)| \exp(-i \phi) \simeq  0.211 \exp( -i \phi)$ when the relative phase $\phi$ between the two beams is varied. Note that, at $\phi= 3 \pi/2$, the CPA condition is obtained, whereas strong amplification, which is the signature of the approaching lasing condition, is obtained by changing the relative phase to $\phi= \pi/2$.\par
{\it Conclusions.} We have investigated the lasing and CPA properties of a $\mathcal{PT}$-symmetric microring structure with phase-shifted index and gain gratings, side coupled to a bus waveguide. The analysis shows that for the complex grating in the broken $\mathcal{PT}$ phase the microring/waveguide system can operate simultaneously as a laser and as a CPA device, providing single mode selection and a controllable unidirectional/bidirectional laser emission.  Such results are expected of aiding the design and development of CPA-laser systems and other on-chip synthetic structures to harness the flow of light in non-conventional ways. Although the complex $\mathcal{PT}$ index modulation considered in our system is not easy to direct implementation, in-phase separated real and imaginary index modulations can exactly mimic the original $\mathcal{PT}$ modulation, which are feasible in fabrication as successfully demonstrated in Ref. \cite{g3}.

\newpage

\newpage

\footnotesize {\bf References with full titles}\\
\\
\noindent
1. C.M. Bender, {\it Making sense of non-Hermitian Hamiltonians}, Rep. Prog. Phys. {\bf 70}, 947 (2007).\\
2. C. E. R\"{u}ter, K.G. Makris, R. El-Ganainy, D.N. Christodoulides, M. Segev, and D. Kip, {\it Observation of parityÐtime symmetry in optics}, Nature Phys. {\bf 6}, 192 (2010).\\
3. A. Regensburger, C. Bersch, M.-A. Miri, G. Onishchukov, D.N. Christodoulides, and	U. Peschel, {\it Parity-time synthetic photonic lattices}, Nature {\bf 488}, 167 (2012).\\
4. L. Feng, Y.-L. Xu, W.S. Fegadolli, M.-H. Lu, J.E.B. Oliveira, V.R. Almeida,	 Y.-F. Chen, and A. Scherer, {\it Experimental demonstration of a unidirectional reflectionless parity-time metamaterial at optical frequencies}, Nature Mat. {\bf 12}, 108 (2013).\\
5. S. Longhi, {\it $\mathcal{PT}$ symmetric laser-absorber}, Phys. Rev. A {\bf 82}, 031801 (2010).\\
6. Y.D. Chong, L. Ge, and A.D. Stone, {\it $\mathcal{PT}$-symmetry breaking and laser-absorber modes in optical scattering systems}, Phys. Rev. Lett. {\bf 106}, 093902 (2011).\\
7. A. Mostafazadeh, {\it Self-dual spectral singularities and coherent perfect absorbing lasers without $\mathcal{PT}$}, J. Phys. A {\bf 45}, 444024 (2012).\\
8. C. Y. Huang, R. Zhang, J. L. Han, J. Zheng, and J. Q. Xu, {\it Type-II perfect absorption and amplification modes with controllable bandwidth in combined $\mathcal{PT}$-symmetric and conventional Bragg-grating structures}, Phys. Rev. A {\bf 89}, 023842 (2014).\\
9. J. Schindler, Z. Lin, J.M. Lee, H. Ramezani, F.M. Ellis, and T. Kottos, 
{\it PT-symmetric electronics}, J. Phys. A {\bf 45}, 444029 (2012).\\
10. Y. Sun, W. Tan, H.-q. Li, J. Li, and H. Chen, {\it Experimental Demonstration of a Coherent Perfect Absorber with PT Phase Transition},
Phys. Rev. Lett. {\bf 112}, 143903 (2014).\\
11. L. Poladian, {\it Resonance mode expansions and exact solutions for nonuniform gratings}, Phys. Rev. E  {\bf 54}, 2963 (1996).\\
12. M. Kulishov, J. M. Laniel, N. Belanger, J. Azana, and D. V. Plant, {\it Nonreciprocal waveguide Bragg gratings}, Opt. Express {\bf 13}, 3068 (2005).\\
13. M. Greenberg and M. Orenstein, {\it Irreversible coupling by use of dissipative optics}, Opt. Lett. {\bf 29}, 451 (2004).\\
14. M. Kulishov, J. M. Laniel, N. Belanger, and D. V. Plant, {\it Trapping light in a ring resonator using a grating-assisted
coupler with asymmetric transmission}, Opt. Express {\bf 13}, 3567 (2005).\\
15. S. Longhi, {\it Optical Realization of Relativistic Non-Hermitian Quantum Mechanics}, Phys. Rev. Lett. {\bf 105}, 013903 (2010).\\
16. Z. Lin, H. Ramezani, T. Eichelkraut, T. Kottos, H. Cao, and D.N. Christodoulides, {\it Unidirectional Invisibility Induced by $\mathcal{PT}$-Symmetric Periodic Structures},
Phys. Rev. Lett. {\bf 106}, 213901 (2011).\\
17. S. Longhi, {\it Invisibility in $\mathcal{PT}$-symmetric complex crystals}, J. Phys. A {\bf 44}, 485302 (2011).\\
18. M. Kulishov and B. Kress, {\it Distributed Bragg reflector structures based on $\mathcal{PT}$-symmetric coupling with lowest possible lasing threshold}, Opt. Express {\bf 21}, 22327 (2013).\\
19. H. Schomerus, {\it Quantum Noise and Self-Sustained Radiation of $\mathcal{PT}$-Symmetric Systems}, Phys. Rev. Lett. {\bf 104}, 233601 (2010).\\
20. G. Yoo, H.-S. Sim, and H. Schomerus, {\it Quantum noise and mode nonorthogonality in non-Hermitian $\mathcal{PT}$-symmetric optical resonators}, Phys. Rev. A {\bf 84}, 063833 (2011).\\
21. V. Van, T.A. Ibrahim, P.P. Absil, F.G. Johnson, R. Grover, P.-T. Ho, {\it Optical Signal Processing Using Nonlinear Semiconductor Microring Resonators}, IEEE
J. Sel. Top. Quantum Electron.  {\bf 8}, 705 (2002).\\
22. M.T. Hill, H.J.S. Dorren, T. de Vries, X.J.M. Leitjens,
J.H. den Besten, B. Samlbrugge, Y.-S. Oei, H. Binsma,
G.-D. Khoe, M.K. Smit, {\it A fast low-power optical memory based on coupled micro-ring lasers}, Nature {\bf 432}, 206 (2004).\\
23. K.J. Vahala, {\it Optical microcavities}, Nature {\bf 424}, 839 (2003)
24. F. Nazari, N. Bender, H. Ramezani, M. K.Moravvej-Farshi, D. N. Christodoulides, and T. Kottos, {\it Optical isolation via $\mathcal{PT}$-symmetric nonlinear Fano resonances}, Opt. Express {\bf 22}, 9575 (2014).\\
25. B. Peng, S. K. Ozdemir, F. Lei, F. Monifi, M. Gianfreda, G. L. Long, S. Fan, F. Nori, C. M. Bender, and L.
Yang, {\it Parity-time-symmetric whispering-gallery microcavities}, Nat. Phys. {\bf 10}, 394 (2014).\\
26. M. Brandstetter, M. Liertzer, C. Deutsch, P. Klang, J. Schšberl, H. E. T\"{u}reci, G. Strasser, K. Unterrainer,
and S. Rotter, {\it Reversing the Pump-Dependence of a Laser at an Exceptional Point}, Nat. Comm. {\bf 5}, 4034 (2014).\\
27. L. Feng, Z.J. Wong, R. Ma, Y. Wang, and X. Zhang, {\it Parity-time synthetic lasers}, arXiv:1405.2863 (2014).\\
28. A. Yariv, {\it Universal relations for coupling of optical power between microresonators and
dielectric waveguides}, Electron. Lett. {\bf 36}, 321 (2000).\\
29. M. Cai, O. Painter, and K. J. Vahala, {\it Observation of critical coupling in a fiber taper to a silica-microsphere
whispering-gallery mode system}, Phys. Rev. Lett. {\bf 85}, 74 (2000).\\
30. H. Ramezani, S. Kalish, I. Vitebskiy, and T. Kottos , {\it Unidirectional Lasing Emerging from Frozen Light in Non-Reciprocal Cavities}, Phys. Rev. Lett. {\bf 112}, 043904 (2014).\\
31. J.M. Lee, S. Factor, Z. Lin, I. Vitebskiy, F.M. Ellis, and T. Kottos, {\it Reconfigurable Directional Lasing Modes in Cavities with Generalized PT Symmetry},
Phys. Rev. Lett. {\bf 112}, 253902 (2014).

\end{document}